\documentclass[a4paper,12pt]{article}
\usepackage{hyperref}
\usepackage[nosort]{cite}

\usepackage{csquotes}
\usepackage{amssymb,amsmath,amsfonts,amsthm,fullpage,epic,eepic,float}
\usepackage[mathscr]{eucal}
\usepackage{amsmath}
\usepackage{amsfonts}
\usepackage{amssymb}
\usepackage{graphicx}
\usepackage{array}
\usepackage{cancel}
\usepackage{caption}
\usepackage{wrapfig}
\usepackage{secdot}
\usepackage{indentfirst}
\usepackage{mathrsfs}
\usepackage{cancel}
\usepackage{misccorr}
\usepackage{amsthm}
\usepackage[left=1cm,right=1cm,top=2cm,bottom=2cm,bindingoffset=0cm]{geometry}
\usepackage[OT1, T2A]{fontenc}
\usepackage[english]{babel}
\usepackage{rotating,indentfirst,array,varioref}
\usepackage{appendix,marginnote,tikz,pgf,mathtools}
\usepackage{setspace}
\usepackage{authblk}
\usepackage{tikz,pgf}

\def\be{\begin{equation}}
\def\ee{\end{equation}}
\def\ba{\begin{aligned}}
\def\ea{\end{aligned}}

\def\barr{\begin{array}}
\def\earr{\end{array}}

\def\ili{\int\limits}
\def\<{\left(}
\def\>{\right)}

\def\l|{\left|}
\def\r|{\right|}

\def\p{\partial}

\begin{document}
\title{Coincidences between Calabi-Yau manifolds of \\Berglund-Hubsch type and Batyrev polytopes}

\author{Mikhail Belakovskiy$^{1,2}$\thanks{belakovskiy.myu@phystech.edu}}
\author{Alexander Belavin$^{1,2,3}$\thanks{belavin@itp.ac.ru} }

\affil{$^1$ L.D. Landau Institute for Theoretical Physics\\
Akademika Semenova av. 1-A\\ Chernogolovka, 142432 Moscow region, Russia}

\affil{$^2$ Moscow Institute of Physics and Technology\\
Dolgoprudnyi, 141700 Moscow region, Russia}

\affil{$^3$ Institute for Information Transmission Problems\\
Bolshoy Karetny per. 19, build.1, Moscow 127051 Russia}

\maketitle
\abstract{ In this article, we consider the phenomenon of complete coincidence of the key properties of pairs of Calabi-Yau manifolds realized as hypersurfaces in two different weighted projective spaces.
More precisely, the first manifold in such a pair is realized as a hypersurface in a weighted projective space, and the second as a hypersurface in the orbifold of another weighted projective space.
The two manifolds in each pair have the same Hodge numbers and special K\"ahler geometry on the  complex structure moduli space  and are associated with the same $N=2$ gauge linear sigma model.
We give the explanation of this interesting coincidence using the  Batyrev's correspondence between  Calabi-Yau manifolds and the reflexive polyhedra.}
\flushbottom 
\section{Introduction}
The set of Calabi-Yau (CY)  manifolds  that can be used as the classical background for compactification in superstring theory includes the set considered by Berglund-Hubsch  (BH list) \cite{BerHub1, BerHub2} which  consists of varieties defined in the weighted projective spaces $P^4_ {(k_1, k_2, k_3, k_4, k_5)}$ as the zero locus of quasi-homogeneous polynomials $ W (x_i) $ belonging to one of $16$ types.

The reason for our interest in the BH list is the observation of complete  coincidence of key properties in some pairs of CY manifolds defined in two different weighted projective spaces from this set.

A similar fact of the coincidence was considered in \cite {COK}. 
In  this case  two polynomials defining the CY families in two different weighted projective spaces are connected by a birational transformation.  In our examples, such transformations do not exist.
and we explain the fact of coincidence in another way.\\
We found two cases of the coincidences. In both cases the first of CY manifolds in a pair is defined as a hypersurface in a weighted projective space,  and the second as an orbifold in another weighted projective space. In addition to the coincidence of the Hodge numbers, as we have checked, in each of  these two cases both CY manifolds have the same special K\"ahler geometry on the spaces of complex modules.
Also we  check the mirror version of the JKLMR-conjecture \cite {Jockers} about the relationship between the K\"ahler potential on moduli space of CY manifold  and the corresponding partition function of  $N=2$ GLSM \cite{Witten:1993yc} for both cases. 

The coincidence of GLSM models, which correspond to two CY's in each of the two pairs, shows that the mirror partners of each of these CY manifolds  also coincide.

In section \ref{Geometry} of this paper, we recall the  method for computing the K\"ahler potential on the of complex structure moduli  space developed in \cite{AKBA1, AKBA2, AKBA3}. In section \ref{coincidence}  we apply this method to compute  K\"ahler potentials for each CY manifold  in both  pairs  and find  the coincidence mentioned above. In section \ref{coincidence-GLSM} we recall the method  for building the GLSM models that corresponds to CY manifold \cite{ABL1, ABL2}.  Using this method we find GLSMs, and they turn out to be the same for both CYs in each pair. Finally, in section \ref{Polytopes} we give an explanation of the matches found in previous sections. For this, we use the correspondence proposed by Batyrev between the CY families and reflexive polyhedra  \cite{Batyrev:1994hm}. We build Batyrev polyhedra for each  CY family  following to the approach of \cite{ABL1} and see that the polyhedra in each pair are the same.
\section{How to compute  geometry on the  complex structure moduli space}\label{Geometry}
In this section we remind the technique of the computation  \cite{AKBA1, AKBA2, AKBA3} of K\"ahler potential of the complex structure moduli space of Calabi-Yau manifolds $X$ defined as  hypersurfaces in weighted projective spaces and their orbifolds.
This technique is based on the deep connection, found in \cite{LEVAWA}, between the cohomology  ring of CY manifold  defined as the zero-locus of a polynomial $W(x)$ of degree $d=\sum\limits_{i=1}^5 k_i$ and the chiral ring $R^Q$ defined by the same polynomial  $W(x)$.

As it is known \cite{COGP,Candelas} on every CY manifold  there exists one non-vanishing holomorphic $(3,0)$ -form $\Omega$. The K\"ahler potential on the complex structure moduli space is defined as 
\begin{equation}
e^{-K_c ^X}=\int\limits_X\Omega\wedge\bar{\Omega}.
\end{equation}
We consider  CY manifolds $X$ that  can be defined as hypersurfaces in the  weighted projective spaces 
\be
P^{4}_{(k_1, \ldots, k_5)} = \{(x_1 : \cdots, x_5) \; | \; (x_1 : \cdots, x_5) \simeq ( \lambda^{k_1} x_1 : \cdots, \lambda^{k_5} x_5)\}.
\ee
These hypersurfaces are the zero locus of a sum of one of the  transverse polynomials BH $W_0$ and $h (= \dim H ^ {2,1})$ monomials, that correspond to the deformations of the complex structure of $X$
\begin{equation}
\begin{aligned}
W(x)=&W_0(x)+\sum\limits_{s=1}^h\phi_{s}e_{s}=0,\\
W(\lambda^{k_i}x_i)&=\lambda^dW(x_i), \;\;\;\;\; d=\sum\limits_{i=1}^5k_i.
\end{aligned}
\end{equation}

The holomorphic $(3,0)$-form in this case can be  explicitly written  in terms of projective coordinates on $P^4_{k_1, \cdots, k_5}$ as 
\begin{equation}
\begin{aligned}
\Omega = \frac{x_5 d x_1\wedge d x_2\wedge d x_{3}}{\p W(x)/\p x_{4}}.
\end{aligned}
\end{equation}
Here  the form  $\Omega$ is restricted  on the zero-locus of  $W$. 

Let  $q_a$ be some basis  in the middle homology group $H_3(X, \mathbb{ R})$, then the K\"ahler potential can be rewritten in terms of periods $\omega_{a}$ over the cycles $q_a$ and their intersection matrix $C_{ab}=q_a\cap q_b$
\be
\ba
e^{-K_c(X)} &= \omega_{a}(\phi) C_{ab}\; \overline{\omega_{b}(\phi)},\\
&\omega_{a}(\phi) := \int_{q_{a}} \Omega.
\ea
\ee

Now we consider the chiral ring $R^Q$, a subring of the Milnor ring $R$ associated with the polynomial $W_0$ 
\begin{equation}
\begin{aligned}
R=&\frac{\mathbb{C}[x_1,...,x_5]}{\langle\p W_0(x)/\p x_i\rangle}.
\end{aligned}
\end{equation}
The chiral ring  $R^Q$ consists of those  elements $P(x_i)$ of the Milnor ring  $R$  that are invariant under action of the group of the "quantum symmetry" $Q=Z_d$
\begin{equation}
\begin{aligned}
P(\omega^{k_i}x_i)&=W(x_i), \;\;\;\;  \omega^d=1.
\end{aligned}
\end{equation}
The ring $R^Q$ has a grading. It can be decomposed in a direct sum of four components which consist of polynomials of degree $0$, $d$, $2d$ and $3d$ with respect to the weights of the  projective coordinates $x_i$. This grading corresponds to the Hodge structure on X:
\be
\ba
R^Q=(R^Q)^0\oplus(R^Q)^1\oplus(R^Q)^2\oplus(R^Q)^3\\
H^3=H^{3,0}\oplus H^{2,1}\oplus H^{1,2}\oplus H^{0,3}.
\ea
\ee

Let the set of  monomials $e_\mu$   be the basis  in the chiral ring $R^Q$ and $\eta_{\mu\nu}$ is  the invariant pairing in it  defined as
\begin{equation}
\eta_{\mu\nu}=Res\frac{e_\mu e_\nu d^5x}{\prod\limits_{i=1}^5\p W_0/\p x_i}.
\end{equation}
The next step is to introduce two differentials
\begin{equation}
\begin{aligned}
D_{\pm}=d\pm dW_0\wedge
\end{aligned}
\end{equation}
and to define  two corresponding to them cohomology groups $H^5_{D_\pm}$.  The groups $H^5_{D_\pm}$ are isomorphic to $R^Q$
and the  set of $5$-forms $e_\mu d^5x$ can be chosen as the basis in each of them. 

We also define two relative homology groups $H_5(\mathbb{C}^5, ReW_0\to\pm\infty)$ dual to $ H ^ 5_ {D_ \pm} $ relative to the pairing
\be
\ba
\langle e_\mu d^5x, Q^\pm_a\rangle=\int\limits_{Q^\pm_a} e_\mu e^{\mp W_0(x)}d^5x
\ea
\ee
Here  the cycles $Q^\pm_a$ belong to the homology groups  $H_5(\mathbb{C}^5, ReW_0\to\pm\infty)$ with real coefficients and determine the bases in them.

The key fact for the approach of \cite{AKBA1} is that  the groups $H_{5, D_\pm}$ and $H_3(X)$ are isomorphic. This isomorphism is established by the following requirement: the  cycle  $q_a \in H_3(X, \mathbb{R})$ is in one to one correspondence  to the cycle  $Q_a \in H_5(\mathbb{C}^5, ReW_0\to\pm\infty)$ iff
\begin{equation}
\int\limits_{q_a}\Omega=\int\limits_{Q_a^\pm}e^{\mp W(x)}d^5x.
\end{equation}
It follows that the intersection matrices in these  two different homology  groups coincide
\begin{equation}
\begin{aligned}
C_{ab}= q_a\cap q_b=Q^+_{a}\cap Q^{-}_{b}.
\end{aligned}
\end{equation}
Also we emphasize that $Q_a^+\cap Q_b^+=Q_a^-\cap Q_b^-=0$.\\

Below we use the important assertion established in \cite{Chiodo} about the connection between the intersection matrix $C_{ab}$ and the pairing $\eta_{\mu \nu}$  in the chiral ring $R^Q$
\begin{equation}
\eta_{\mu\nu}=\langle e_\mu, e_\nu\rangle=
\int\limits_{Q^+_a}e_\mu e^{-W_0}d^5x C_{ab}\int\limits_{Q^-_b}e_\nu e^{W_0}d^5x.
\end{equation}
Introducing  the  matrix $T^\pm$ 
\begin{equation}
\begin{aligned}
T_{a\mu}^\pm:=\int\limits_{Q^\pm_a}e_\mu e^{\mp W_0}d^5xa,
\end{aligned}
\end{equation}
 we can rewrite this relation as
\begin{equation}
\begin{aligned}
\eta_{\mu \nu}=T^+_{a\mu}C_{ab}T^-_{b\nu}.
\end{aligned}
\end{equation}
Expressing from  this relation  the intersection matrix $C_{ab}$ in terms the pairing $\eta_{\mu \nu}$ we can rewrite the formula for the K\"ahler potential
\be
\ba
e^{-K_c(X)} &= \omega_{a}(\phi) C_{ab}\; \overline{\omega_{b}(\phi)},\\
\ea
\ee
as follows
\begin{equation}
\ba
e^{-K_c(X)}=\sigma^+_\mu\eta_{\mu\lambda}M_{\lambda\nu}\overline{\sigma^-_\nu}
\ea
\end{equation}
where  $M:=T^{-1}\overline{T}$ 
and  $\sigma_\mu^{\pm}:=T^{-1}_{\mu a} \omega_a $ and
$$\omega_{a}(\phi) := \int_{q_{a}} \Omega=\int\limits_{Q_a^\pm}e^{\mp W(x)}d^5x.$$ The resulting  formula for K\"ahler potential is very useful for the computation.

Firstly, instead of the unknown intersection matrix $ C_ {ab} $, it uses the pairing in the chiral ring $ R ^ Q $, which can be easily calculated.

Secondly, if we introduce a convenient basis cycles  $\Gamma^\pm_\mu$  in the  relative homology groups    
$H_5(\mathbb{C}^5, ReW_0\to\pm\infty)$  dual to the chosen basis in $H^5_{D_\pm}$ as
\begin{equation}
\begin{aligned}
\int\limits_{\Gamma_\mu^\pm}e_\nu e^{\mp W_0}d^5x=\delta_{ \mu, \nu},
\end{aligned}
\end{equation}
then we can easily check  that the matrix  $T_{a \mu }$, which is  a transition matrix 
from a real basis $Q_{a}$ to the complex basis $\Gamma_\mu$, connects the periods  as follows 
\begin{equation}
\begin{aligned}
\omega_{a}=T^{\pm}_{a\mu}\sigma^{\pm}_\mu.
\end{aligned}
\end{equation}
and that  the matrix  $M:=T^{-1}\overline{T}$  does not depend on the choice of the real basis $Q_{a}$. 
 Indeed, if another basis $\{\tilde{Q}_a\}=S_{ab}Q_b$, where $S$ is a real matrix, is chosen, then $\tilde{T}_{a\mu}=S_{ab}T_{b\mu}$ and the new matrix $$\tilde{M}=\tilde{T}^{-1}\bar{\tilde{T}}=T^{-1}S^{-1}\bar{S}\bar{T}=T^{-1}\bar{T}=M$$.

This allows us to choose a real basis $Q_ {a}$ in such a way that the matrices $T ^ {\pm}$ are easily computable. The example of such a choice are the Lefschetz thimbles \cite{AKBA3}.

Thirdly,  we find that the  periods $\sigma^\pm_\mu=\int\limits_{\Gamma^\pm_\mu}e^{\mp W(x)}d^5x$
 defined  as the integrals over the cycles  $\Gamma^\pm_\mu$, can be efficiently computed.
Namely, the integrals for  periods $\sigma_\mu=\int\limits_{\Gamma^\pm_\mu}e^{\mp W(x)}d^5x$ can be computed by decomposing the exponent by parameters $\phi_a$
\begin{equation}
\begin{aligned}\label{periods}
\sigma_\mu^{\pm}=  \int\limits_{\Gamma^\pm_\mu}e^{\mp W(x)}d^5x&=
\sum\limits_{m_1,...m_{h^{2,1}}=1}^{\infty}
 C^\pm_{\{\mu| m_1,\cdots,m_h\}}
 \prod\limits_{s=1}^{h^{2,1}}\frac{\phi_s^{m_s}}{m_s!},\\
 C^\pm_{\{\mu| m_1,\cdots,m_h\}}=&\int\limits_{\Gamma^\pm_\mu}e^{\mp W_0(x)} \prod\limits_{s=1}^{h^{2,1}} e_s^{m_s}d^5x.
\end{aligned}
\end{equation}
Each term $ \int\limits_{\Gamma^\pm_\mu}e^{\mp W_0(x)} \prod\limits_{s=1}^{h^{2,1}} e_s^{m_s}d^5x$
 in this formula depends only on the  $H^5_{D_\pm}$  cohomology class of the monomial. Therefore it can be simplified using  the relation 
\begin{equation}
\begin{aligned}
P d^5x= P' d^5x+D_{\pm}\Psi,
\end{aligned}
\end{equation}
where $\Psi$ is  a suitable 4-form. 
Due to the relation   $\int\limits_{\Gamma^\pm_\mu}e^{\mp W_0(x)}  P d^5x=\int\limits_{\Gamma^\pm_\mu} e^{\mp W_0(x)}  P'd^5x $  the degree of the monomial  $P$ can be reduced.
Repeating this procedure, we obtain explicit expression for the  periods $\sigma_\mu$.
\section{Coincidence of the special geometry on  the  moduli spaces of pairs of CY families}\label{coincidence}
In this section, we consider two pairs  of Calabi-Yau manifolds. In each of these pairs  there are two differently defined manifolds, however,  the K\"ahler potentials on their complex structure moduli spaces  coincide, as it follows from their comparison.                
\subsection{The first pair of Calabi-Yau manifolds}
Here we consider  two Calabi-Yau  families. We call them  1a and 1b. 
They are defined by different types of BH polynomials in different projective spaces. 
But as we show the explicit computation of the special K\"ahler geometry gives the same result in both cases.
\paragraph{The case 1a.}
This Calabi-Yau family $X$ in this case is defined as a hypersurface in $P^4_{23, 55, 28, 53, 34}$  by the equation 
\begin{equation}
\begin{aligned}
&W_0+\phi_1e_1+\phi_2e_2=0\\
&W_0=x_1^6x_2+x_2^3x_3+x_3^5x_4+x_4^3x_5+x_5^5x_1.
\end{aligned}
\end{equation}
The  basis in the chiral ring $R^Q$ is given by  six monomials
\begin{equation}
e_0=1,\; e_1=x_1x_2x_3x_4x_5,\; e_2=x_1^3x_3^2x_5^2,\; e_3=e_2^2,\; e_4=e_2e_3,\; e_5=e_2^2e_3.
\end{equation}
The Hodge numbers of $X$ are $h^{2,1}=2, h^{1,1}=95$. 
The   $e_1$ and $e_2$ define the complete set of deformations of the complex structure.

Following the approach, presented in the section $2$, we  simplify the expression for the periods \eqref{periods}. 
For this,  we use the fact that  two 5-forms $P_1d^5x$ and $P_2d^5x$   belong to the same cohomology class,  $P_1d^5x \sim P_2d^5x$,   if  $P_1d^5x-P_2d^5x=D_{\pm}\Psi$.
 
After some calculations we obtain the five equivalence relations
\begin{equation}
\begin{aligned}
\prod\limits_{j=1}^5 x_j^{a_j} d^5x \sim (1-B_{ij}(a_j+1)) \prod_{j=1}^5 x_j^{a_j-M_{ij}}d^5x,       
i=1, ..., 5,
\end{aligned}
\end{equation}
where $W_0:=\sum\limits_{i=1}^5\prod\limits_{j=1}^5x_i^{M_{ij}}$ and  $B_{ij}:=(M)^{-1}_{ij}$.
Then each monomial in \eqref{periods} can be replaced by another using these five relations:
\begin{equation}
\begin{aligned}
&e_1^ne_2^md^5x\sim -\frac{1}{7}(n+3m-6)e_1^{n-1}e_2^{m-2}d^5x,\\
&e_1^ne_2^md^5x\sim \frac{1}{7}(2n-m-5)e_1^{n-3}e_2^{m+1}d^5x.
\end{aligned}
\end{equation}
Applying these two formulas step by step each monomial in \eqref{periods} can be reduced to one of six monomials from the chiral ring according to the following relation:
\begin{equation}
\begin{aligned}
e_1^ne_2^md^5x\sim& (-1)^m\frac{\Gamma^3(\frac{n+3m+1}{7})}{\Gamma^3(\frac{\mu}{7})}\frac{\Gamma^2(\frac{2n-m+2}{7})}{\Gamma^2(\frac{\nu}{7})}e_\mu d^5x,\\
                       \mu=&n+3m-6 (\textrm{mod}\, 7), 1\leqslant \mu \leqslant 6, \\
                       \nu=&2\mu (\textrm{mod}\, 7), 1\leqslant \nu \leqslant 6.
\end{aligned}
\end{equation}
Then taking into account the definition of the cycles $\Gamma^\pm_\mu$ we obtain the following result for the periods over these cycles
\begin{equation}
\begin{aligned}
\sigma_\mu(\phi_1, \phi_2)=\int\limits_{\Gamma_\mu}e^{-W(x)}(x)d^5x=\sum\limits_{n,m}\int\limits_{\Gamma_\mu}e^{-W_0(x)}e_1^ne_2^m\frac{\phi_1^n\phi_2^m}{n!m!}d^5x=\\=\sum\limits_{n,m\in\Sigma_{\mu}}(-1)^m\frac{\Gamma^3(\frac{n+3m+1}{7})}{\Gamma^3(\frac{\mu}{7})}\frac{\Gamma^2(\frac{2n-m+2}{7})}{\Gamma^2(\frac{\nu}{7})}\frac{\phi_1^n\phi_2^m}{n!m!},
\end{aligned}
\end{equation}
where
\begin{equation}
\begin{aligned}
\Sigma_{{\mu}}=\{n,m\in \mathbb{N}_0|n+3m-6=\mu(mod 7), 0\leqslant\mu\leqslant6\}, \nu=\nu(\mu)=2\mu(mod7), 0\leqslant\nu\leqslant 6.
\end{aligned}
\end{equation}

In order to find  the K\"ahler potential it is still necessary  to know the transition matrix $T^\pm_{a\mu}$ from the cycles $\Gamma^\pm_\mu$ to some real cycles $Q^\pm_a$.\\
 Before  computing $T_{a\mu}$ it is convenient  to change the variables as
$$
x_i=\prod\limits_{j=1}^5 y_j^{7 B_{ij}},
$$
such that in the new variables $y_i$
\begin{equation}
\begin{aligned}
W_0(x(y))=\sum\limits_{i=1}^5y_i^7.
\end{aligned}
\end{equation}
Now we have a Fermat polynomial and it is possible to choose a basis of real cycles as described in \cite{AKBA3}, where the  special Lefschetz thimbles are used as real cycles. The only difference is that we need to remember about the Jacobian of the transformation $J=det\frac{\p x}{\p y}$. After the change of variables we have:
\begin{equation}
\begin{aligned}
T_{a\mu}=\int\limits_{L_a}e_\mu(x(y))e^{-W_0(x(y))}J(y)d^5y
\end{aligned}
\end{equation}
Knowing  the $T_{a\mu}$ we compute the  matrix $M=T^{-1}\overline{T}$ and obtain
\begin{equation}
\begin{aligned}
M_{\mu\nu}=\delta_{\mu,6-\nu}\gamma^3(\frac{\mu}{7})\gamma^2(\frac{\delta}{7})
\end{aligned}
\end{equation}
where $\delta(\mu)=2\mu(mod7)$, $1\leqslant\delta\leqslant6$.

Now the result for the K\"ahler potential on the complex moduli space is obtained:
\begin{equation}
\begin{aligned}
e^{-K_c(\phi_1, \phi_2)}=\sum\limits_{\mu=1}^{6}\gamma^3(\frac{\mu}{7})\gamma^2(\frac{\nu}{7})|\sigma_{\mu}(\phi_1, \phi_2)|^2
\end{aligned}
\end{equation}
where $\nu(\mu)=2\mu(mod7)$, $1\leqslant\nu\leqslant6$.
\paragraph{The case 1b.}This Calabi-Yau family is defined as a hypersurface in  the orbifold of the weighted projective space $P^4_{3,2,7,2,7}/\mathbb{Z}_{21}$  by the equation 
\begin{equation}
\begin{aligned}
&W_0+\phi_1e_1+\phi_2e_2=0\\
&W_0=x_1^7+x_2^7x_3+x_3^3+x_4^7x_5+x_5^3.
\end{aligned}
\end{equation}
The complete basis in the chiral ring is given by six monomials
\begin{equation}
\begin{aligned}
e_0=1,\; e_1=x_1x_2x_3x_4x_5,\; e_2=x_1^3x_2^3x_4^3,\; e_3=e_2^2,\; e_4=e_2e_3,\; e_5=e_2^2e_3.
\end{aligned}
\end{equation}
Its Hodge numbers are $h^{2,1}=2, h^{1,1}=95$. There $e_1$ and $e_2$ define the complete set of deformations of the complex structure.

The special K\"ahler geometry in this case was computed in \cite{AKBA4}. The result  is as follows
\begin{equation}
\begin{aligned}
e^{-K_c(\phi_1, \phi_2)}=\sum\limits_{\mu=1}^{6}\gamma^3(\frac{\mu}{7})\gamma^2(\frac{\nu}{7})|\sigma_{\mu}(\phi_1, \phi_2)|^2,
\end{aligned}
\end{equation}
where $\nu(\mu)=2\mu(mod7)$, $1\leqslant\nu\leqslant6$.

As we see, the K\"ahler potential in this case  completely coincides with that obtained above for the case 1a. Thus the two differently defined families 1a and 1b  have the identical special K\"ahler geometry. This fact gives a hint of the equivalence of these two families.
\subsection{The second pair}
Here we introduce the second couple of CY families (we call them  2a and 2b). And here it also turns out that these two families have the same special K\"ahler geometry.
\paragraph{The case 2a.} This Calabi-Yau family is defined as a hypersurface in $P[4]_{(97, 53, 35, 37, 25)}$ by the equation
\begin{equation}
\begin{aligned}
&W_0+\phi_1e_1+\phi_2e_2=0,\\
&W_0=x_1^2x_2+x_2^4x_3+x_3^6x_4+x_4^6x_5+x_5^6x_1.
\end{aligned}
\end{equation}
The basis in the chiral ring $ R ^ Q $ consists of the following six monomials
\begin{equation}
\begin{aligned}
e_0=1,\; e_1=x_1x_2x_3x_4x_5,\;e_2=x_2x_3^2x_4^2x_5^2,\; e_3=e_2^2,\; e_4=e_1e_2,\; e_5=e_1e_2^2.
\end{aligned}
\end{equation}
The Hodge numbers in this case are $h^{2,1}=2, h^{1,1,}=122$.  The monomials  $e_1$ and $e_2$ define the  set of the complex structure deformations. After computations similar to those performed in the case 1a, we obtain the expression for the K\"ahler potential
\begin{equation}
e^{-K_c(\phi_1, \phi_2)}=\sum\limits_{\mu=1}^{6}\gamma(\frac{\mu}{7})\gamma^4(\frac{\nu}{7})|\sigma_{\mu}(\phi_1, \phi_2)|^2,
\end{equation}
where $\nu=3\mu (\textrm{mod}\, 7), \; 1 \leqslant \nu \leqslant 6$.
\paragraph{The case 2b.} This Calabi-Yau family  is defined in $P^4_{7,2,2,2,1}/\mathbb{Z}_{7}^2$ by the equation
\begin{equation}
\begin{aligned}
&W(x)=W_0+\phi_1e_1+\phi_2e_2=0\\
&W_0=x_1^2+x_2^7+x_3^7+x_4^7+x_5^7x_1
\end{aligned}
\end{equation}
The complete basis in the chiral ring is given by six monomials:
\begin{equation}
e_0=1,\; e_1=x_1x_2x_3x_4x_5,\;e_2=x_2^2x_3^2x_4^2x_5^2,\; e_3=e_2^2,\; e_4=e_1e_2,\; e_5=e_1e_2^2.
\end{equation}
Though it is not a loop polynomial the computation of the special geometry does not deviate from the previous cases. Proceeding the same computations we find the expressions of periods which completely coincide with the expression for the previous family
\begin{equation}
\begin{aligned}
\sigma_\mu(\phi_1, \phi_2)=\sum\limits_{n,m\in\Sigma_{{\mu}}}(-1)^n\frac{\Gamma(\frac{3n-m+3}{7})}{\Gamma(\frac{\mu}{7})}\frac{\Gamma^4(\frac{n+2m+1}{7})}{\Gamma^4(\frac{\nu}{7})}\frac{\phi_1^n\phi_2^m}{n!m!},
\end{aligned}
\end{equation}
\begin{equation}
\begin{aligned}
\Sigma_{{\mu}}=\{n,m\in \mathbb{N}_0|3n-m-4=\mu(mod 7)\}, \nu=-2\mu(mod 7), 0\leqslant\nu\leqslant6\}.
\end{aligned}
\end{equation}
And the final result for the K\"ahler potential on the complex moduli structures space is
\begin{equation}
\begin{aligned}
e^{-K_c(\phi_1, \phi_2)}=\sum\limits_{\mu=1}^{6}\gamma(\frac{\mu}{7})\gamma^4(\frac{\nu}{7})|\sigma_{\mu}(\phi_1, \phi_2)|^2,
\end{aligned}
\end{equation}
where $\nu=3\mu (mod 7), \; 1 \leqslant \nu \leqslant 6$.

As we see, it completely coincides with obtained above for the case 2a. Thus the two differently defined families 2a and 2b  have the identical special K\"ahler geometry. Therefore, as in the previous case, we can expect the equivalence of these two families.
\section{Coincidence of   GLSM's, corresponding to both CY families, in each of the pairs}\label{coincidence-GLSM}
In this part, we first recall the construction of \cite{ABL1, ABL2} of the  $N =2$  GLSM model, which corresponds to  a given CY manifold defined as a hypersurface in some weighted projective spase. 
We find  that CY manifolds of the cases 1a and 1b, as well as the 2a and 2b,  correspond to the same GLSM model. Also in this section, we verify the JKLMR-conjecture \cite{Jockers}.

The mirror version of the JKLMR-conjecture \cite{BonTan, Gomis1, Gomis2,Gomis3} states that the two-sphere  partition function $Z_Y$ of the GLSM model, computed exactly by localization in \cite{BeniniCremonesi, Doroud}, gives  the exact K\"ahler potential on the  moduli space  of complex structures for the  Calabi-Yau manifold $X$, which  is the mirror of the vacuum manifold $Y$ of the GLSM
\begin{equation}
Z_Y=e^{-K_C^X}.
\end{equation}

We build the GLSM   using the fan associated with CY family $Y$. 
It is known \cite{Mirror} that the vectors of the  skeleton of the  fan for the CY family $Y$, correspond to the vertices of the  Batyrev's polyhedron for the CY family $X$,  if $X$ and $Y$ are mirrors. Therefore, starting  with the polynomial $W_X$ and the Batyrev's polyhedron which corresponds to it ,  we find  a fan of the mirror CY manifold $Y$ from it  and use this knowledge to  construct  the corresponding GLSM for $Y$ , as was done  in \cite{ABL1, ABL2, Eremin}.
 
We find the polyhedron corresponding to the family $X$, determined by this polynomial, as follows.  We rewrite the polynomial  $W_X=W_0+\sum\limits_{l=1}^h\phi_l e_a$ in the following form
\begin{equation}
W=\sum\limits_{a=1}^5\prod\limits_{j=1}^5x_i^{V_{aj}}+
\sum\limits_{a=6}^{h+5}\prod\limits_{j=1}^5\phi_s x_j^{V_{aj}}.
\end{equation}
Here we have $5\times(h+5)$-matrix $V_{ai}$ or, equivalently, $h+5$ five-dimensional lattice vectors $V_a$.  In fact,  they lie in the $4$-dimensional sublattice, defined by the equation
\begin{equation}
 \sum\limits_{i=1}^5 V_{ai} k_i=d.
\end{equation}
Batyrev's  4-dimensional reflexive  polyhedron is defined as a convex hull of points $V_{ai}$.  The lattice points $ V_ {ai}$ of this polyhedron are vectors that form the edges of the fan for the mirror family  $Y $ \cite{COK}. Since the fan of the family Y consists of $ 5 + h $ five-dimensional vectors, between them there exist $ h $ linear independent relations
\begin{equation}
\begin{aligned}
Q_{la}V_{ai}=0, \;\; l=\overline{1,h},\;\;a=\overline{1,h+5},\;\;i=\overline{1,5}.
\end{aligned}
\end{equation}
We  have to find the such a set $Q_{la}$ that define an integral  basis in the space of the linear relations  between  the vectors $V_{ai}$. As a result, we obtain $ h $ sets of charges of $ 5 + h $ chiral fields for the $ h $ $ U (1) $ gauge groups
which we use to determine the corresponding GLSM.
\subsection{Cases 1a and 1b}
The test of JKLMR hypothesis for these two  cases can easily  be reduced  to a single computation. For the family 1a the points of the Batyrev polyhedron have the following coordinates
\begin{equation}
\begin{aligned}
v_1:\;\;\;
&\begin{pmatrix}
6,&1,&0,&0,&0
\end{pmatrix}\\
v_2:\;\;\;
&\begin{pmatrix}
0,&3,&1,&0,&0
\end{pmatrix}\\
v_3:\;\;\;
&\begin{pmatrix}
0,&0,&5,&1,&0
\end{pmatrix}\\
v_4:\;\;\;
&\begin{pmatrix}
0,&0,&0,&3,&1
\end{pmatrix}\\
v_5:\;\;\;
&\begin{pmatrix}
1,&0,&0,&0,&5
\end{pmatrix}\\
v_6:\;\;\;
&\begin{pmatrix}
1,&1,&1,&1,&1
\end{pmatrix}\\
v_7:\;\;\;
&\begin{pmatrix}
3;&0;&2;&0;&2
\end{pmatrix}.
\end{aligned}
\end{equation}
As for the 1b family, they are following
\begin{equation}
\begin{aligned}
v_1:\;\;\;
&\begin{pmatrix}
7,&0,&0,&0,&0
\end{pmatrix}\\
v_2:\;\;\;
&\begin{pmatrix}
0,&7,&1,&0,&0
\end{pmatrix}\\
v_3:\;\;\;
&\begin{pmatrix}
0,&0,&3,&0,&0
\end{pmatrix}\\
v_4:\;\;\;
&\begin{pmatrix}
0,&0,&0,&7,&1
\end{pmatrix}\\
v_5:\;\;\;
&\begin{pmatrix}
0,&0,&0,&0,&3
\end{pmatrix}\\
v_6:\;\;\;
&\begin{pmatrix}
1,&1,&1,&1,&1
\end{pmatrix}\\
v_7:\;\;\;
&\begin{pmatrix}
3,&3,&0,&3,&0
\end{pmatrix}.
\end{aligned}
\end{equation} 

Between these seven five-dimensional vectors, there must be two linearly independent relations with integer coefficients.
We have to  choose the integral basis of the linear  relations. 
That is, to choose two such integer coefficients of linear relations $ Q_ {1a}, Q_ {2a} $ such that the coefficients of  any integral  relation $ Q'_a $ between the vectors $ v_a $ can be represented as:
\begin{equation}
\begin{aligned}
Q'_a=m_1Q_{1a}+m_2Q_{2a},
\end{aligned}
\end{equation}
where $m_1,m_2\in\mathbb{Z}$.

We call $Q_{1a}, Q_{2a}$ a  $\mathbb{Z}$-basis in the space of the linear integer relations. 
It is easy to see that the $\mathbb{Z}$-basis of these relations can be chosen similarly in both these cases:
\begin{equation}
\begin{aligned}
Q_{la}v_a=0; l=1,2; a=1,2,3,4,5,
\end{aligned}
\end{equation}
where
\begin{equation}
\begin{aligned}
Q_{1a}=
\begin{pmatrix}
1,0,1,0,1,-1,-2
\end{pmatrix},
Q_{2a}=
\begin{pmatrix}
0,1,0,1,0,-3,1
\end{pmatrix}.
\end{aligned}
\end{equation}

As a result  we get  two sets of the charges of $5$ chiral fields   for the two $U(1)$ gauge groups of  the corresponding GLSM. 
Also  we choose the R-symmetry charges \cite {BeniniCremonesi, Doroud, Eremin} as
\begin{equation}
\begin{aligned}
q_1=q_3=q_5=\frac{2}{7}\\
q_2=q_4=\frac{4}{7}\\
q_6=q_7=0.
\end{aligned}
\end{equation}
Then the partition function of this sigma-model can be presented as the following expression dependent on four real Fayet-Illiopoulos parameters \cite{BeniniCremonesi, Doroud}
\begin{equation}
\begin{aligned}
Z(r_1, r_2, \theta_1, \theta_2)=\sum\limits_{m_1, m_2 \in \mathbb{Z}}e^{-i\theta_1m_1-i\theta_2m_2}\ili\ili\frac{d\tau_1}{2\pi i}\frac{d\tau_2}{2\pi i}e^{4\pi r_1\tau_1+4\pi r_2\tau_2}\times\\ \times \prod\limits_{a=1}^{7}\frac{\Gamma(\frac{q_a}{2}+Q_{1a}(\tau_1-\frac{m_1}{2})+Q_{1a}(\tau_2-\frac{m_2}{2}))}{\Gamma(1-\frac{q_a}{2}-Q_{1a}(\tau_1+\frac{m_1}{2})-Q_{1a}(\tau_2+\frac{m_2}{2}))}.
\end{aligned}
\end{equation}
Now changing variables allows one to express four real parameters with two complex variables
\begin{equation}
\begin{aligned}
&z_1=\exp\left(-\frac{2\pi}{7}r_1-\frac{4\pi}{7}r_2-\frac{4\pi i}{7}\theta_1-\frac{8\pi i}{7}\theta_2\right)\\
&z_2=\exp\left(-\frac{6\pi}{7}r_1+\frac{2\pi}{7}r_2-\frac{12\pi i}{7}\theta_1+\frac{4\pi i}{7}\theta_2\right).
\end{aligned}
\end{equation}
Or more precisely
\begin{equation}
\begin{aligned}
&\;\;\;\;\;\;\;\;Z=\sum\limits_{m_1\in\mathbb{Z}}\sum\limits_{m_2\in\mathbb{Z}}\ili\ili\frac{d\tau_1}{2\pi i}\frac{d\tau_2}{2\pi i}
\frac{\Gamma^3(\frac{1}{7}+(\tau_1-\frac{m_1}{2}))}{\Gamma^3(\frac{6}{7}-(\tau_1-\frac{m_1}{2}))}
\frac{\Gamma^2(\frac{2}{7}+(\tau_2-\frac{m_2}{2}))}{\Gamma^2(\frac{5}{7}-(\tau_2+\frac{m_2}{2}))}\times \\
&\;\;\;\;\;\;\;\;\times\frac{\Gamma(-(\tau_1-\frac{m_1}{2})-3(\tau_2-\frac{m_2}{2}))}{\Gamma(1+(\tau_1+\frac{m_1}{2})+3(\tau_2+\frac{m_2}{2}))}
\frac{\Gamma(-2(\tau_1-\frac{m_1}{2})+(\tau_2-\frac{m_2}{2}))}{\Gamma(1+2(\tau_1+\frac{m_1}{2})-(\tau_2+\frac{m_2}{2}))}\times \\
&\times z_1^{-(\tau_1-\frac{m_1}{2})-3(\tau_2-\frac{m_2}{2})}\bar{z}_1^{-(\tau_1+\frac{m_1}{2})-3(\tau_2+\frac{m_2}{2})}
z_2^{-2(\tau_1-\frac{m_1}{2})+(\tau_2-\frac{m_2}{2})}\bar{z}_2^{-2(\tau_1+\frac{m_1}{2})(\tau_2+\frac{m_2}{2})}.
\end{aligned}
\end{equation}

We compute this integral by  the Cauchy theorem and  obtain
\begin{equation}
\begin{aligned}
&Z=\sum\limits_{{\mu}}\sum\limits_{(n,m),(\bar{n}, \bar{m})\in\Sigma_{{\mu}}}z_1^{-n}z_2^{-m}\bar{z}_1^{-\bar{n}}\bar{z}_2^{-\bar{m}}\frac{\Gamma^3(\frac{1+n+3m}{7})}{\Gamma^3(1-\frac{1+\bar{n}+3\bar{m}}{7})}\frac{\Gamma^2(\frac{2+2n-m}{7})}{\Gamma^2(1-\frac{2+2\bar{n}-\bar{m}}{7})}\frac{(-1)^{n+m}}{n!m!\bar{n}!\bar{m}!}=\\ &=\sum\limits_{{\mu}}\sum\limits_{(n,m),(\bar{n}, \bar{m})\in\Sigma_{{\mu}}}\frac{z_1^{-n}z_2^{-m}\bar{z}_1^{-\bar{n}}\bar{z}_2^{-\bar{m}}}{n!m!\bar{n}!\bar{m}!}sin^3(\frac{\pi}{7}(1+\bar{n}+3\bar{m}))sin^2(\frac{\pi}{7}(1+2\bar{n}-\bar{m}))\times \\&\times \Gamma^3(\frac{1+n+3m}{7})\Gamma^2(\frac{2+2n-m}{7})\Gamma^3(\frac{1+\bar{n}+3\bar{m}}{7})\Gamma^2(\frac{2+2\bar{n}-\bar{m}}{7})= \\ &\sum\limits_{{\mu}}\gamma^3(\frac{\nu}{7})\gamma^2(\frac{\mu}{7})\sum\limits_{(n,m),(\bar{n}, \bar{m})\in\Sigma_{{\mu}}}(-1)^{m+n+\bar{m}+\bar{n}}z_1^{-n}z_2^{-m}\bar{z}_1^{-\bar{n}}\bar{z}_2^{-\bar{m}}\gamma^3(\frac{\nu}{7})\gamma^2(\frac{\mu}{7})\times \\ & \times\Gamma^3(\frac{1+n+3m}{7})\Gamma^3(\frac{1+\bar{n}+3\bar{m}}{7})\Gamma^3(\frac{2+2n-m}{7})\Gamma^3(\frac{2+2\bar{n}-\bar{m}}{7})=\\&=\sum\limits_{{\mu}}\gamma^3(\frac{\nu}{7})\gamma^2(\frac{\mu}{7})|\sigma_{{\mu}}(-\frac{1}{z_1}, \frac{1}{z_2})|^2.
\end{aligned}
\end{equation}
This can be rewritten as
\begin{equation}
Z =\sum\limits_{{\mu}}\gamma^3(\frac{\nu}{7})
\gamma^2(\frac{\mu}{7})|\sigma_{{\mu}}(-\frac{1}{z_1}, \frac{1}{z_2})|^2
\end{equation}
and after  the change of variables, which connects coordinates on the complex moduli structures space and Fayet-Illiopoulos parameters
\begin{equation}
\phi_1=-\frac{1}{z_1},\quad
\phi_2=\frac{1}{z_2},
\end{equation}
exactly coincides with the expression for $e^{-K_c(\phi_1, \phi_2)}$ obtained above. 
 Here  $K_c(\phi_1, \phi_2)$  is the K\"ahler potential on the complex moduli spaces for the both CY manifolds of the first pair.
\subsection{Cases 2a and 2b}
In the second  pair, we again have two sets of the lattice points.
The seven points in the case 2a
\begin{equation}
\begin{aligned}
v_1:\;\;\;
\begin{pmatrix}
2,&1,&0,&0,&0
\end{pmatrix}\\
v_2:\;\;\;
\begin{pmatrix}
0,&4,&1,&0,&0
\end{pmatrix}\\
v_3:\;\;\;
\begin{pmatrix}
0,&0,&6,&1,&0
\end{pmatrix}\\
v_4:\;\;\;
\begin{pmatrix}
0,&0,&0,&6,&1
\end{pmatrix}\\
v_5:\;\;\;
\begin{pmatrix}
1,&0,&0,&0,&6
\end{pmatrix}\\
v_6:\;\;\;
\begin{pmatrix}
1,&1,&1,&1,&1
\end{pmatrix}\\
v_7:\;\;\;
\begin{pmatrix}
0,&1,&2,&2,&2
\end{pmatrix}
\end{aligned}
\end{equation}
and the seven points in 2b-case
\begin{equation}
\begin{aligned}
v_1:\;\;\;
\begin{pmatrix}
2,&0,&0,&0,&0
\end{pmatrix}\\
v_2:\;\;\;
\begin{pmatrix}
0,&7,&0,&0,&0
\end{pmatrix}\\
v_3:\;\;\;
\begin{pmatrix}
0,&0,&7,&0,&0
\end{pmatrix}\\
v_4:\;\;\;
\begin{pmatrix}
0,&0,&0,&7,&0
\end{pmatrix}\\
v_5:\;\;\;
\begin{pmatrix}
1,&0,&0,&0,&7
\end{pmatrix}\\
v_6:\;\;\;
\begin{pmatrix}
1,&1,&1,&1,&1
\end{pmatrix}\\
v_7:\;\;\;
\begin{pmatrix}
0,&2,&2,&2,&2
\end{pmatrix}.
\end{aligned}
\end{equation}

We again find  that the integral basis of the linear  relations can be chosen the same in both of these cases
\begin{equation}
\begin{aligned}
Q_{la}v_a=0; l=1,2; \quad a=1,2,3,4,5,
\end{aligned}
\end{equation}
where
\begin{equation}
\begin{aligned}
Q_{2a}=
\begin{pmatrix}
1,0,0,0,0,-2,1
\end{pmatrix},\quad
Q_{1a}=
\begin{pmatrix}
0,1,1,1,1,-1,-3
\end{pmatrix}.
\end{aligned}
\end{equation}
We choose the $R$-symmetry charges 
\begin{equation}
q_1=q_2=q_3=q_4=\frac{2}{7},\quad
q_5=\frac{6}{7},\quad
q_6=q_7=0
\end{equation}
and  change the  variables as follows
\begin{equation}
\begin{aligned}
&z_2=\exp\left(-\frac{4\pi}{7}r_2+\frac{2\pi}{7}r_1-\frac{8\pi i}{7}\theta_2+\frac{4\pi i}{7}\theta_1\right)\\
&z_1=\exp\left(-\frac{2\pi}{7}r_2-\frac{6\pi}{7}r_1-\frac{4\pi i}{7}\theta_2-\frac{12\pi i}{7}\theta_1\right).
\end{aligned}
\end{equation}
After that,  for the partition function we get 
\begin{equation}
\begin{aligned}
&\;\;\;\;\;\;\;\;Z=\sum\limits_{m_1\in\mathbb{Z}}\sum\limits_{m_2\in\mathbb{Z}}\ili\ili\frac{d\tau_1}{2\pi i}\frac{d\tau_2}{2\pi i}
\frac{\Gamma^4(\frac{1}{7}+(\tau_1-\frac{m_1}{2}))}{\Gamma^4(\frac{6}{7}-(\tau_1-\frac{m_1}{2}))}
\frac{\Gamma(\frac{3}{7}+(\tau_2-\frac{m_2}{2}))}{\Gamma(\frac{4}{7}-(\tau_2+\frac{m_2}{2}))}\times\\
&\;\;\;\;\;\;\;\;\times\frac{\Gamma(-2(\tau_1-\frac{m_1}{2})+(\tau_2-\frac{m_2}{2}))}{\Gamma(1+2(\tau_1+\frac{m_1}{2})-(\tau_2+\frac{m_2}{2}))}
\frac{\Gamma(-(\tau_1-\frac{m_1}{2})+3(\tau_2-\frac{m_2}{2}))}{\Gamma(1+(\tau_1+\frac{m_1}{2})-3(\tau_2+\frac{m_2}{2}))}\times \\
&\times z_1^{-3(\tau_1-\frac{m_1}{2})+(\tau_2-\frac{m_2}{2})}\bar{z}_1^{-3(\tau_1+\frac{m_1}{2})+(\tau_2+\frac{m_2}{2})}
z_2^{-(\tau_1-\frac{m_1}{2})-2(\tau_2-\frac{m_2}{2})}\bar{z}_2^{-(\tau_1+\frac{m_1}{2})-2(\tau_2+\frac{m_2}{2})}.
\end{aligned}
\end{equation}
Taking the integral we get  the following expression
\begin{equation}
\begin{aligned}
Z=\sum\limits_{\mu=1}^6\gamma^4(\frac{\mu}{7})\gamma(\frac{\nu}{7})|\sigma_{\mu}(\frac{1}{z_1}, -\frac{1}{z_2})|^2.
\end{aligned}
\end{equation}
At last after  changing the variables  
\begin{equation}
\phi_1=z_1^{-1},\quad
\phi_2=-z_2^{-1},
\end{equation}
this expression again 
exactly coincides with the expression for $e^{-K_c(\phi_1, \phi_2)}$ obtained above for the K\"ahler potential on the complex moduli space for the both CY manifolds of the second pair.

\section{ Batyrev's polytopes and the coincidences of the CY families}\label{Polytopes}

In this section we are going  to explain the found coincidence of  the properties of the CY families in the first and the second pairs of CY families using the Batyrev approach \cite {Batyrev:1994hm}. We construct reflexive polyhedra corresponding to each family CY in the first pair, and it turns out that families 1a and 1b define the same polyhedron. This is also true for the second pair.  That explains  the coincidence of the CY families in both cases.

When constructing reflexive polyhedra, we follow the scheme presented in the previous chapter, with a single modification. We know that the vectors $v_a$,  where $a=1,...,h_{21}+5$,  forming the Batyrev polyhedron obey the following relations
\begin{equation}
\begin{aligned}
\sum\limits_{i=1}^5k_iv_{ai}=\sum\limits_{i=1}^5k_i:=d.
\end{aligned}
\end{equation}
That is they lie in the same hyperplane, where  the end of  the vector $ v_ \rho $ is located
\begin{equation}
\begin{aligned}
v_\rho=\begin{pmatrix}
1\\1\\1\\1\\1
\end{pmatrix}.
\end{aligned}
\end{equation}
After shifting all vectors to the vector $ v_ \rho $: $ \tilde {v} _a = v_a-v_ \rho $ the polyhedron falls into a hyperplane that contains the origin. Later in this section we will omit the tilde symbol.

The reflexive polyhedron $ \Delta $ corresponds to the polynomial $ W_X $ and is formed as the convex hull of the exponents $ v_a $ of the monomials that make up $ W_X $. If the CY family $ X $ is defined as a hypersurface in a weighted projective space, then the vectors $ v_a $   can be written in the form of the linear combinations with the  integer coefficients of the vectors of the integral basis of the four-dimensional lattice $ M$ \cite{Mirror}.  The lattice $ M$ defined by the equation $\sum\limits_{i=1}^5m_i k_i=0$ , where   
$(m_1,m_2 m_3, m_4, m_5) $ are the integer points   belonging to the five-dimensional lattice $\mathbb{Z}^5$.

If the CY  family is defined as a hypersurface in a quotient  by a discrete group $H$ of some weighted projective space,  then the vectors $ v_a $   should  be written in the form of linear combinations with integer coefficients of the vectors of the integral basis of the sublattice  $ M/H $ of the lattice $M$.
Below we demonstrate the difference between these two situation.

To make sure that a certain set of vectors forms a basis in the lattice $M$ that is the  sublattice of the  $\mathbb{Z}^5$, defined  by the equation
\begin{equation}
\begin{aligned}
\sum\limits_{i=1}^5k_im_i=0,
\end{aligned}
\end{equation}
where  the $k_i$ are coprime numbers, we  use the  following criterion.

The four linearly independent vectors $u_1,u_2,u_3,u_4 \in M$ constitute a basis in $M$ if and only if the volume $\textrm{Vol}_{\textrm{cell}}$ of the cell generated by the vectors $u_1,u_2,u_3,u_4$ equals to minimal possible non-zero volume of the cell in $M$ which is equal to $|k|=\sqrt{k_1^2+k_2^2+k_3^2+k_4^2+k_5^2}$. This volume is equal to  the module  of the vector $V$
\begin{equation}
\begin{aligned}
\textrm{Vol}_{\textrm{cell}}=\sqrt{V_1^2+V_2^2+V_3^2+V_4^2+V_5^2},
\end{aligned}
\end{equation}
if   five components of the vector  $V$ defined as follows
\begin{equation}
\begin{aligned}
V_j=\epsilon_{ji_1i_2i_3i_4}u_{1i_1}u_{2i_2}u_{3i_3}u_{4i_4}.\\
\end{aligned}
\end{equation}
Since the vector $k$ with components $k_i$ is perpendicular to any vector of  the lattice $M$ and  the vector $V$, the criterion can be reformulated as follows: \\  
the   four vectors $u_1,u_2,u_3,u_4$   constitute an integral  basis in $M$ if and only if 
 the volume $Vol(k,u_1,...u_{4})$ generated by them and by the fifth  "vector" $k$ with the components $k_i$ satisfies  the relation
\begin{equation}
Vol(k,u_1,...u_{4})=|k|^2.
\end{equation}
\subsection{Cases 1a and 1b}
Consider the first pair of  Calabi-Yau families. The first of them, 1a, is a hypersurface in $P^4_{23,55,28,53,34}$, given by the equation 
$$W_1(x)=x_1^6x_2+x_2^3x_3+x_3^5x_4+x_4^3x_5+x_5^5x_1+\phi_1x_1x_2x_3x_4x_5+\phi_2x_1^3x_3^2x_5^2=0$$. \\
 The second is given by the equation 
 $$W_2(x)=x_1^7+x_2^7x_3+x_3^3+x_4^7x_5+x_5^3+\psi_1x_1x_2x_3x_4x_5+\psi_2x_1^3x_2^3x_4^3=0$$
 in $P^4_{3,2,7,2,7}/\mathbb{Z}_{21}$, where the action of $\mathbb{Z}_{21}$  is defined by the formula
\begin{equation}
(x_1,x_2,x_3,x_4,x_5)\to(\omega^{12}x_1, \omega^{2}x_2, \omega^7x_3, x_4,x_5)
\end{equation}
and  $\omega^{21}=1$.

To construct  Batyrev polyhedron, corresponding to the family 1a, we take the lattice $M$.  It is given by the equation $23m_1+55m_2+28m_3+53m_4+34m_5=0$ in $\mathbb{Z}^5$ with $m_i$ . 
The lattice points  of the polyhedron in $\mathbb{Z}^5$ are
\begin{equation}
\begin{aligned}
f_1:\;\;\;
&\begin{pmatrix}
&5,&0,-1,-1,-1
\end{pmatrix}\\
f_2:\;\;\;
&\begin{pmatrix}
-1,&2,&0,-1,-1
\end{pmatrix}\\
f_3:\;\;\;
&\begin{pmatrix}
-1,-1,&4,&0,-1
\end{pmatrix}\\
f_4:\;\;\;
&\begin{pmatrix}
-1,-1,-1,&2,&0
\end{pmatrix}\\
f_5:\;\;\;
&\begin{pmatrix}
&0,-1,-1,-1,&4
\end{pmatrix}\\
f_6:\;\;\;
&\begin{pmatrix}
&2,-1,&1,-1,&1
\end{pmatrix}.
\end{aligned}
\end{equation}
If we try to  choose the integral basis $\{e_i\}_{i=1}^4$ in M as follows
\begin{equation}
e_i=f_i, i=1,2,3,4,
\end{equation}
or more explicitly as
\begin{equation}
\begin{aligned}
e_1:\;\;\;
&\begin{pmatrix}
&5,&0,-1,-1,-1
\end{pmatrix}\\
e_2:\;\;\;
&\begin{pmatrix}
-1,&2,&0,-1,-1
\end{pmatrix}\\
e_3:\;\;\;
&\begin{pmatrix}
-1,-1,&4,&0,-1
\end{pmatrix}\\
e_4:\;\;\;
&\begin{pmatrix}
-1,-1,-1,&2,&0,
\end{pmatrix}
\end{aligned}
\end{equation}
then we find that this basis satisfies  the above criterion.
The lattice points $v_i$ of the polyhedron in this basis are expressed as follows
\begin{equation}
\begin{aligned}
v_1:\;\;\;
&\begin{pmatrix}
&1,&0,&0,&0
\end{pmatrix}\\
v_2:\;\;\;
&\begin{pmatrix}
&0,&1,&0,&0
\end{pmatrix}\\
v_3:\;\;\;
&\begin{pmatrix}
&0,&0,&1,&0
\end{pmatrix}\\
v_4:\;\;\;
&\begin{pmatrix}
&0,&0,&0,&1
\end{pmatrix}\\
v_5:\;\;\;
&\begin{pmatrix}
-1,-2,-1,-2
\end{pmatrix}\\
v_6:\;\;\;
&\begin{pmatrix}
&0,-1,&0,-1
\end{pmatrix}
\end{aligned}
\end{equation}

The convex hull of these six vertices $\{v_i\}_{i=1}^6$ determines the  polyhedron of the family 1a.  It is verified that the just built polyhedron is reflexive. Because it has the only point in its interior of this polyhedron - the origin  and the lattice points $m_i$  in its facets are subject of the equations
\begin{equation}
\begin{aligned}
\sum\limits_{i=1}^5(n_i m_i)=1,
\end{aligned}
\end{equation}
where $n_i$ is a vector with four integral coprime components.

Now let us construct the polyhedron for the family 1b.In this case we have  the lattice $N$ that is given by the equation $3m_1+2m_2+7m_3+2m_4+7m_5=0$ in $\mathbb{Z}^5$. 
The lattice points  of the polyhedron in $N$ are as follows
\begin{equation}
\begin{aligned}
f'_1:\;\;\;
&\begin{pmatrix}
&6,-1,-1,-1,-1
\end{pmatrix}\\
f'_2:\;\;\;
&\begin{pmatrix}
-1,-1,&2,-1,-1
\end{pmatrix}\\
f'_3:\;\;\;
&\begin{pmatrix}
-1,&6,&0,-1,-1
\end{pmatrix}\\
f'_4:\;\;\;
&\begin{pmatrix}
-1,-1,-1,-1,&2
\end{pmatrix}\\
f'_5:\;\;\;
&\begin{pmatrix}
-1,-1,-1,&6,&0
\end{pmatrix}\\
f'_6:\;\;\;
&\begin{pmatrix}
&2,&2,-1,&2,-1
\end{pmatrix}.
\end{aligned}
\end{equation}
If here, as in case 1a, the family was defined by hyperspace in the weighted projective space, and not in its orbifold, then the construction could be completed on this step.
But it would turn out that the corresponding  polyhedron is not reflexive. 
Since, although there is a single point in the interior of this polyhedron, 
its faces cannot be determined by the equations
\begin{equation}
\begin{aligned}
\sum\limits_{i=1}^5(n_i m_i)=1,
\end{aligned}
\end{equation}
where $n_i$ is a vector with four integral coprime components.


Returning to the process of constructing a polyhedron of this family, we can answer the question of why it leads to a non-reflexive polyhedron. The fact is that it is necessary to take into account the difference between how families 1a and 1b are determined. 
The second is defined in the quotient of the projective space by the group $ \mathbb {Z} _ {21} $.
This also needs to be considered when constructing the polyhedron.
In fact, the polyhedron must be defined in the sublattice of the lattice $N$.
This sublattice $N/\mathbb{Z}_{21}$ is defined by the equations
\begin{equation}
\begin{aligned}
&3m_1+2m_2+7m_3+2m_4+7m_5=0\\
&12m_1+2m_2+7m_3=0, \mod {21}.
\end{aligned}
\end{equation}
To construct the reflexive  polyhedron of family 1b, it is necessary to choose the integral basis in the lattice $N/\mathbb{Z}_{21}$.
We verify that the set of four vectors $ \{f'_1, f'_2, f'_3, f'_4 \} $ is an integral basis in the sublattice
 $N/\mathbb{Z}_{21}$.

First, each vector from the set  $ \{f'_1, f'_2, f'_3, f'_4 \} $
 satisfies the relation
\begin{equation}
\begin{aligned}
12(f'_1)_i+2(f'_2)_i+7(f'_3)_i=0, \mod {21}.
\end{aligned}
\end{equation}
It is also necessary to verify that any vector of the sublattice  $ N / \mathbb {Z} _ {21} $
can be represented as the sum of vectors $ \{f'_i\} _{i = 1}^ 4 $ with integer coefficients.\\
To do this, it is enough to show that the cell volume generated by the four vectors $ \{f'_1, f'_2, f'_3, f'_4 \} $ from the $ N / \mathbb {Z} _ {21} $ sublattice is exactly $21$ times larger
than the volume of the cell formed by the basis of the lattice $ N $ itself.
This can be done as explained above
\begin{equation}
\begin{aligned}
Vol(k, e_1, e_2,e_3,e_4)=\begin{pmatrix}
&3&2&7&2&7\\
&6&-1&-1&-1&-1\\
&-1&-1&2&-1&-1\\
&-1&6&0&6&0\\
&-1&-1&-1&-1&2
\end{pmatrix}=21*|k|^2=21*115.
\end{aligned}
\end{equation}
Therefore, indeed, $ \{f'_1, f'_2, f'_3, f'_4 \} $ is an integral basis in this sublattice.

In this basis in $N/\mathbb{Z}_{21}$ the lattice points  of the polyhedron have the following coordinates
\begin{equation}
\begin{aligned}
v'_1:\;\;\;
&\begin{pmatrix}
&1,&0,&0,&0
\end{pmatrix}\\
v'_2:\;\;\;
&\begin{pmatrix}
&0,&1,&0,&0
\end{pmatrix}\\
v'_3:\;\;\;
&\begin{pmatrix}
&0,&0,&1,&0
\end{pmatrix}\\
v'_4:\;\;\;
&\begin{pmatrix}
&0,&0,&0,&1
\end{pmatrix}\\
v'_5:\;\;\;
&\begin{pmatrix}
-1,-2,-1,-2
\end{pmatrix}\\
v'_6:\;\;\;
&\begin{pmatrix}
&0,-1,&0,-1
\end{pmatrix}.
\end{aligned}
\end{equation}
Thus, polyhedra of the families 1a and 1b indeed coincide.
\subsection{Cases 2a и 2b}
In this part also two Calabi-Yau families are considered. The first is given in $P^4_{97,53,35,37,25}$ by the equation 
$$W_1(x)=x_1^2x_2+x_2^4x_3+x_3^6x_4+x_3^6x_5+x_6^5x_1+
\phi_1x_1x_2x_3x_4x_5+\phi_2x_2x_3^2x_4^2x_5^2=0.$$ The second is given  in $P^4_{3,2,7,2,7}/\mathbb{Z}_{7}^2$ by the equation 
$$W_2(x)=x_1^7+x_2^7x+x_3^7+x_4^7+x_5^2x_1+\psi_1x_1x_2x_3x_4x_5+
\psi_2x_2^2x_3^2x_4^2x_5^2=0,$$
 where the action of $\mathbb{Z}_{7}^2$ is defined  as
\begin{equation}
\ba
(x_1,x_2,x_3,x_4,x_5)\to(x_1, \omega_1^{-1}x_2, \omega_1 x_3, \omega_2^{-1}x_4,\omega_2x_5).
\ea
\ee
The polyhedron of the family 2a is built like a polyhedron of the family 1a:
\begin{equation}
\begin{aligned}
f_1:\;\;\;
&\begin{pmatrix}
&1,&0,-1,-1,-1
\end{pmatrix}\\
f_2:\;\;\;
&\begin{pmatrix}
-1,&3,&0,-1,-1
\end{pmatrix}\\
f_3:\;\;\;
&\begin{pmatrix}
-1,-1,&5,&0,-1
\end{pmatrix}\\
f_4:\;\;\;
&\begin{pmatrix}
-1,-1,-1,&5,&0
\end{pmatrix}\\
f_5:\;\;\;
&\begin{pmatrix}
&0,-1,-1,-1,&5
\end{pmatrix}\\
f_6:\;\;\;
&\begin{pmatrix}
-1,&0,&1,&1,&1
\end{pmatrix}.
\end{aligned}
\end{equation}
The set of four vectors $\{e_i\}_{i=1}^4, e_i=f_i, i=1,..,4$ is a basis in the lattice $M=\{(m_1,...,m_5)|97m_1+53m_2+35m_3+37m_4+25m_5=0\}$. Then the polyhedron has the following lattice points  in this basis
\begin{equation}
\begin{aligned}
v_1:\;\;\;
&\begin{pmatrix}
&1,&0,&0,&0
\end{pmatrix}\\
v_2:\;\;\;
&\begin{pmatrix}
&0,&1,&0,&0
\end{pmatrix}\\
v_3:\;\;\;
&\begin{pmatrix}
&0,&0,&1,&0
\end{pmatrix}\\
v_4:\;\;\;
&\begin{pmatrix}
&0,&0,&0,&1
\end{pmatrix}\\
v_5:\;\;\;
&\begin{pmatrix}
-1,-1,-1,-3
\end{pmatrix}\\
v_6:\;\;\;
&\begin{pmatrix}
&0,&0,&0,-1
\end{pmatrix}.
\end{aligned}
\end{equation}
In the case 2b  the  family is defined in the orbifold $P^4_{7,2,2,2,1}/\mathbb{Z}_7^2$ and  the $\mathbb{Z}_7\times\mathbb{Z}_7$-action on the weighted projective space is given by the formula
\begin{equation}
\begin{aligned}
(x_1,x_2,x_3,x_4,x_5)&\to(x_1, \omega_1^{-1}x_2, \omega_1 x_3, \omega_2^{-1}x_4,\omega_2x_5),\\
\omega_1^7&=\omega_2^7=1.
\end{aligned}
\end{equation}
Now the construction is similar to case 1b. To build the polyhedron it is necessary to consider the lattice $N=\{m\in\mathbb{Z}^5|7m_1+2m_2+2m_3+2m_4+m_5=0\}$. 
The lattice points of the polyhedron are as follows
\begin{equation}
\begin{aligned}
f'_1:\;\;\;
&\begin{pmatrix}
&1,-1,-1,-1,-1
\end{pmatrix}\\
f'_2:\;\;\;
&\begin{pmatrix}
-1,&6,-1,-1,-1
\end{pmatrix}\\
f'_3:\;\;\;
&\begin{pmatrix}
-1,-1,&6,-1,-1
\end{pmatrix}\\
f'_4:\;\;\;
&\begin{pmatrix}
-1,-1,-1,&6,-1
\end{pmatrix}\\
f'_5:\;\;\;
&\begin{pmatrix}
&0,-1,-1,-1,&6
\end{pmatrix}\\
f'_6:\;\;\;
&\begin{pmatrix}
-1,&1,&1,&1,&1
\end{pmatrix}.
\end{aligned}
\end{equation}
To construct  the polytope we need to choose a basis in the sublattice $N/\mathbb{Z}_{7}^2$. This sublattice  can be given by equations
\begin{equation}
\begin{aligned}
7m_1+2&m_2+2m_3+2m_4+m_5=0,\\
&m_2-m_3=0  \mod 7,\\
&m_4-m_5=0 \mod 7.
\end{aligned}
\end{equation}
Now we have to find  the set of the  four vectors $\{e_i\}_{i=1}^4$ that  generates the sublattice  $N/\mathbb{Z}_{7}^2$. 

Each vector from the set $\{f'_a\}_{a=1}^6, a=1,...,6$ satisfies two relations:
\begin{equation}
\begin{aligned}
(f'_a)_2-(f'_a)_3=0\,(\textrm{mod}\, 7)\\
(f'_a)_4-(f'_a)_5=0\,(\textrm{mod}\, 7).
\end{aligned}
\end{equation}
It is necessary now to make sure there is no other relation which satisfy these vectors. 
It is possible to show that volume of the cell generated by four vectors $\{f'_1,f'_2,f'_3,f'_4\}$ is precisely 49 times bigger than the volume of the cell which forms a basis in the sublattice $N$.  
It follows from the equation
\begin{equation}
\begin{aligned}
Vol(k, f'_1, f'_2,f'_3,f'_4)=\begin{pmatrix}
&7&2&2&2&1\\
&1&-1&-1&-1&-1\\
&-1&6&-1&-1&-1\\
&-1&-1&6&-1&-1\\
&-1&-1&-1&6&-1
\end{pmatrix}=7^2*|k|^2=49*62.
\end{aligned}
\end{equation}
It means that these four vectors give the integral basis of $N/\mathbb{Z}_{7}^2$.
Choosing this basis in $N/\mathbb{Z}_{7}^2$ we represent the six lattice points  of the polytope in this basis as
\begin{equation}
\begin{aligned}
v'_1:\;\;\;
&\begin{pmatrix}
&1,&0,&0,&0
\end{pmatrix}\\
v'_2:\;\;\;
&\begin{pmatrix}
&0,&1,&0,&0
\end{pmatrix}\\
v'_3:\;\;\;
&\begin{pmatrix}
&0,&0,&1,&0
\end{pmatrix}\\
v'_4:\;\;\;
&\begin{pmatrix}
&0,&0,&0,&1
\end{pmatrix}\\
v'_5:\;\;\;
&\begin{pmatrix}
-1,-1,-1,-3
\end{pmatrix}\\
v'_6:\;\;\;
&\begin{pmatrix}
&0,&0,&0,-1
\end{pmatrix}.
\end{aligned}
\end{equation}
So, once again,  we see that the polyhedra of the families in the pair  coincide. This provides an explanation of the coincidences of the Calabi-Yau families themselves.
\section*{Conclusion}
We examined two cases of coincidence of  special geometries on moduli spaces and GLSM for the  CY manifolds defined differently. We found that Batyrev's reflexive polyhedra for such CY-manifolds also coincide,  which is consistent with the coincidences of the special geometries on moduli spaces.
It would be interesting to understand how many different ways there are to build the same Calabi-Yau family from the Berglund-Hubsch list in a more general case.

\section*{Acknowledgements}
We are grateful to G.~ Koshevoy and A.~ Litvinov for useful discussions. This work has been supported by the Russian Science Foundation under the grant 18-12-00439.


\end{document}